\def\E{\ensuremath{\text{E}}}
\def\ah{\ensuremath{\left| h \right|}}
\def\h2{\ensuremath{\left| h \right|^2}}
\def\X{\ensuremath{\mathcal{X}}}
\def\exp{\ensuremath{\text{exp}}}
\def\I{\ensuremath{\mathcal{I}}}
\def\pper{\ensuremath{\varphi_{\text{per}}}}
\def\phalf{\ensuremath{\frac{\pper}{2}}}
\def\pquart{\ensuremath{\frac{\pper}{4}}}
\def\popt{\ensuremath{\varphi^\ast}}
\def\ephi{\ensuremath{e^{i\varphi}}}
\def\epsi{\ensuremath{e^{i\psi}}}
\def\epsiphi{\ensuremath{e^{i(\psi+\varphi)}}}

\def\hV{\ensuremath{h_{\text{VSI}}}}

\newcommand{\reff}[1]{(\ref{#1})}

\newcounter{MYtempeqncnt} 
\newtheorem{prop}{Proposition}

\documentclass[conference]{IEEEtran}

\ifCLASSINFOpdf
  \usepackage[pdftex]{graphicx}
  \graphicspath{{Figures/}}
  \DeclareGraphicsExtensions{.pdf,.jpeg,.png}
\else
  \usepackage[dvips]{graphicx}
  \graphicspath{{Figures/}}
  \DeclareGraphicsExtensions{.eps}
\fi
\usepackage{url}

\usepackage[pdfpagelabels]{hyperref}

\usepackage{cite}
\usepackage{amsmath, amsfonts}

\usepackage{subfig}

\newlength{\myarraycolsep}
\newlength{\oldarraycolsep}
\begin{document}
%
\title{Achievable Rates in Two-user Interference Channels with Finite Inputs and (Very) Strong Interference}

\author{\IEEEauthorblockN{Frederic Knabe and Aydin Sezgin}
\IEEEauthorblockA{Emmy-Noether Research Group on Wireless Networks\\
Institute of Telecommunications and Applied Information Theory\\
Ulm University, Albert-Einstein-Allee 43, 89075 Ulm, Germany\\
Email: \{frederic.knabe,aydin.sezgin\}@uni-ulm.de}
}


%


\maketitle

\begin{abstract}
For two-user interference channels, the capacity is known for the case where interference is stronger than the desired signal. Moreover, it is known that if the interference is above a certain level, it does not reduce the capacity at all. To achieve this capacity, the channel inputs need to be Gaussian distributed. However, Gaussian signals are continuous and unbounded. Thus, they are not well suited for practical applications. In this paper, we investigate the achievable rates if the channel inputs are restricted to finite constellations. Moreover, we will show by numerical simulations that rotating one of these input alphabets in the complex plane can increase the achievable rate region. Finally, we show that the threshold at which the single-user rates are achieved also depends on this rotation.
\end{abstract}


%
\IEEEpeerreviewmaketitle

\begin{figure*}[!t]
\normalsize
\setcounter{MYtempeqncnt}{\value{equation}}
\setcounter{equation}{5}
\begin{equation}\label{eq:mi_fin_dis}
I(Y_1;X_2)=\log_2(M)-\frac{1}{M^2}\sum_{k=1}^M\sum_{l=1}^M \E \left\{ \log_2 \left( \frac{\sum_{\tilde{k}=1}^M\sum_{\tilde{l}=1}^M \exp (-\left| (x_1^k-x_1^{\tilde{k}})+(\tilde{x}_2^l-\tilde{x}_2^{\tilde{l}})\,h+z_1\right|^2\cdot P/2)}{\sum_{\tilde{k}=1}^M \exp (-\left| x_1^k-x_1^{\tilde{k}}+z_1\right|^2\cdot P/2)}\right) \right\}
\end{equation}
\begin{equation}
\label{eq:mi_fin_nodis}
I(Y_1;X_1|X_2)=\log_2(M)-\frac{1}{M}\sum_{k=1}^M \E \left\{ \log_2 \left( \frac{\sum_{\tilde{k}=1}^M \exp (-\left| x_1^k-x_1^{\tilde{k}}+z_1\right|^2\cdot P/2)}{\exp (-\left| z_1\right|^2\cdot P/2)}\right) \right\}
\end{equation}
\setcounter{equation}{\value{MYtempeqncnt}}
\hrulefill
\vspace*{4pt}
\end{figure*}

\section{Introduction}
In modern communication systems, a receiver usually receives signals coming from different transmitting stations. However, in most cases, the receiver is only interested in the signal from one of the transmitting stations. All other signals are considered as interference, which is one of the most important factor limiting the overall performance. The information theoretic model which reflects this situation is the interference channel (IC), where several transmitter-receiver pairs communicate and mutually disturb each other's receptions. Although the interference channel has been studied for decades, the complete capacity region of this channel is still unknown, even for the simplest case of two interfering transmissions (2-IC). 

For the 2-IC, it has been shown that, surprisingly, interference does not reduce capacity when it is above a certain threshold \cite{Car75}. This case is called ``very strong interference''. For the case of ``strong interference'', where the interference is not very strong but still stronger than the desired signal, the capacity is known, too \cite{Car75,HK81}. For all other cases the best known strategy is due to Han and Kobayashi \cite{HK81}. A simplified Han-Kobayashi scheme was recently been shown to achieve rates which are within one bit of the capacity of the 2-IC \cite{ETW08}. A special case of the 2-IC is obtained if one of the receivers is interference-free. This is referred to as the two-user Z-interference channel (2-Z-IC). Also for this channel, the capacity region is only known for strong and very strong interference \cite{Sat81}. However, in contrast to the 2-IC, the sum-capacity is known for all scenarios \cite{Cos85}. 

In order to achieve the capacity (in those cases where it is known), the channel inputs have to be Gaussian distributed. However, Gaussian alphabets are continuous and unbounded. Thus, they are not well suited for practical applications. Naturally the following question arises: What are the achievable rate pairs if the channel inputs are restricted to finite sets like M-PSK or M-QAM, which are used in practical applications?

In this paper, we consider this question for the strong and very strong interference regime of the 2-IC and the 2-Z-IC. We will show that in these channels, rotation of one input alphabet leads to a substantial gain in achievable sum rate. Moreover, we show that the very strong interference threshold depends on the rotation and channel phases, i.e., by applying rotation the very strong interference regime can be reached earlier. This is in contrast to the noisy interference regime, where the threshold up to which treating interference as noise is optimal depends only on the absolute values of the channel gains for both Gaussian and finite alphabets \cite{CJ09}.

 Interestingly, the impact of rotation was analyzed for the two-user broadcast and multiple-access channel (MAC) in \cite{DR09,HR08} recently. Although the 2-IC can be described as two overlapping MACs and some results from \cite{HR08} apply, finding the optimal rotation is not straight forward, as will be shown later.

The paper is organized as follows: Section \ref{sec:system} introduces the interference channel model as well as its capacity region and the required definitions and formulas. An analysis of the impact of rotation of the alphabets in the complex plane is given in section \ref{sec:rotation}. In section \ref{sec:results} we evaluate the achievable rates in the 2-Z-IC and 2-IC numerically. Finally, section \ref{sec:conclusion} concludes the paper.

\section{System description}\label{sec:system}

\subsection{Channel model}
\begin{figure}[b!]
\centering
    \includegraphics[width=.9\linewidth]{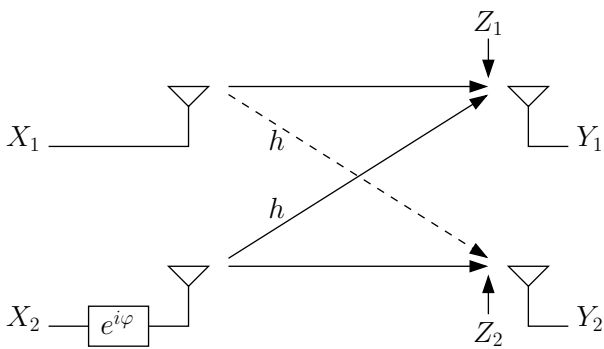}
    \caption{Two-user symmetric (Z-)IC with rotation at transmitter two}
    \label{fig:bothICs}
\vspace{-0.5cm}
\end{figure}

We consider a symmetric interference channel in standard form \cite{Car78} and with complex finite input constellations. Rotation in the complex plane is (w.l.o.g.) applied at transmitter two by multiplying the symbols with $\ephi$. Further rotation is given on the interfering paths, where the path gain is $h=\ah\epsi$. Thus, the received signals are
\begin{equation}\label{eq:2IC}
\begin{bmatrix}Y_1\\Y_2\end{bmatrix} = \begin{bmatrix}1&\ah\epsiphi \\\ah\epsi&\ephi\end{bmatrix}\cdot\begin{bmatrix}X_1\\X_2\end{bmatrix} + \begin{bmatrix}Z_1\\Z_2\end{bmatrix},
\end{equation}
for the case of the 2-IC, where $Z_i\sim\mathcal{CN}(0,1)$ ($i=1,2$) is the Gaussian noise at receiver $i$ and $X_i \in \X_i = \{x_i^1,\ldots,x_i^M\}$ is the channel input at transmitter $i$, which has to fulfill the power constraint $\E\{\left|X_i\right|^2\}\leq P$. Moreover, we set $\X_1=\X_2$, i.e., the transmitters use the same alphabets of size $M$ for transmission. Thus, the received symbols (ignoring the noise) are from  superpositions of two finite sets, where by channel influence one of the sets is a rotated and stretched version of the other one. This stretching and rotation angle between the ``desired'' and the ``interfering'' sets both have an effect on the performance if finite alphabets are used. By the rotation at transmitter two, the rotations between the sets can be influenced and selected such that an optimal performance is obtained.

Both the 2-IC and the 2-Z-IC are visualized in Figure \ref{fig:bothICs}. The 2-Z-IC differs from the 2-IC in that the dashed path from transmitter one to receiver two is not present. Hence, we obtain
\begin{equation}\label{eq:2ZIC}
\begin{bmatrix}Y_1\\Y_2\end{bmatrix} = \begin{bmatrix}1&\ah\epsiphi\\0&\ephi\end{bmatrix}\cdot\begin{bmatrix}X_1\\X_2\end{bmatrix} + \begin{bmatrix}Z_1\\Z_2\end{bmatrix}.
\end{equation}
at the receivers.

\subsection{Capacity region} \label{subsec:capreg}
As already stated in the introduction, we only consider the case of strong or very strong interference, i.e. ($\h2 \geq 1$). For this case the achievable rates $R_1$ and $R_2$ have to satisfy \cite{HK81}
\begin{subequations}\label{eq:SI-Cap}
\begin{align}
 R_1 & \leq I(Y_1;X_1|X_2) \label{eq:mi_R1}\\
 R_2 & \leq I(Y_2;X_2|X_1) \label{eq:mi_R2}\\
 R_1+R_2 & \leq I(Y_1;X_1,X_2)\label{eq:mi_sum_Rx1}\\
 R_1+R_2 & \leq I(Y_2;X_1,X_2)\label{eq:mi_sum_Rx2}.
\end{align}
\end{subequations}
These inequalities have to be fulfilled both for the 2-IC and the 2-Z-IC. However, for the 2-Z-IC \reff{eq:mi_sum_Rx2} is redundant. For maximizing the mutual information terms, circularly symmetric Gaussian input alphabets have to be used. Due to their symmetry, the performance of these alphabets is not subject to rotation by channel or transmitter. The capacity region achieved by Gaussian alphabets $\X_1$ and $\X_2$ can be written as
\begin{subequations}\label{eq:SI-Gauss}
\begin{align}
 R_1 & \leq \log_2(1+P) \label{eq:Gauss_R1}\\
 R_2 & \leq \log_2(1+P) \label{eq:Gauss_R2}\\
 R_1+R_2 & \leq \log_2(1+\h2 P+P). \label{eq:Gauss_Rsum}
\end{align}
\end{subequations}
We obtain only three conditions, because with Gaussian inputs \reff{eq:mi_sum_Rx1} and \reff{eq:mi_sum_Rx2} become equivalent. The last condition is only active if $1\leq \h2 \leq P+1$, which is the strong interference case. For $\h2 \geq P+1$, the IC is in the very strong interference regime and \reff{eq:Gauss_Rsum} becomes redundant.

\subsection{Achievable rates with Finite Constellations}
If finite constellations are used instead of Gaussian alphabets, the terms on the right side of \reff{eq:SI-Cap} have different values. To simplify the evaluation, the chain rule of mutual information
\begin{equation}
I(Y_i;X_i,X_j) = I(Y_i;X_j)+I(Y_i;X_i|X_j)
\end{equation}
is applied. Moreover, we define $\tilde{\X}_2=\{\tilde{x}_2^1,\ldots,\tilde{x}_2^M\}=\X_2\cdot\ephi=\{x_2^1\ephi,\ldots,x_2^M\ephi\}$ as the rotated alphabet of transmitter two. With these definitions, the right side of \reff{eq:SI-Cap} can be expressed with \reff{eq:mi_fin_dis} and \reff{eq:mi_fin_nodis} shown on top of the page, cf. \cite{HR08}. Note that the corresponding terms for receiver two can be obtained by swapping indices $1$ and $2$ for the 2-IC. In the Z-IC it has to be considered that $I(Y_2;X_1)=0$.
\addtocounter{equation}{2}

Of course, there are numerous extensions and further optimizations possible for increasing the achievable rates, such as allowing $\X_1 \neq \X_2$ , using other constellations then QAM, etc.. However, for reasons of simplicity, we restrict ourselves to QAM alphabets here and show that already by allowing rotation, the rate increase is significant. An analysis of the impact of rotation will be given in the next section. 

\section{The impact of rotation}\label{sec:rotation}
If the considered symmetric ICs are in the strong interference regime, the 2-IC can be seen as 2 interlaced multiple-access channels (MAC), while the 2-Z-IC has the same rate constraints as one MAC. Thus, we can refer to some observations of \cite{HR08}, where rotation was considered for increasing the rates with finite alphabets in the MAC. It can be shown that rotating the alphabet of one of the users affects only the expressions $I(Y_1;X_2)$, but not $I(Y_1;X_1|X_2)$. Thus, by rotation we can only influence the constraints on the sum rate \reff{eq:mi_sum_Rx1} and \reff{eq:mi_sum_Rx2}. Since for the 2-IC, rotation influences two terms simultaneously, finding the optimum rotation angle is not straight forward.

Depending on the symmetry of an alphabet $\X$, the rotation of this alphabet has a periodicity $\pper \leq 2\pi$, which can be expressed as
\begin{equation}
 \pper = \min_\varphi \; \{\varphi \,|\, \X \ephi = \X\}.
\end{equation}
Thus, all rotations can be taken modulo $\pper$. For the QAM constellations that we consider here, we have 
\begin{equation}
\varphi_{\text{per}_{\text{QAM}}} = \pi/2.
\end{equation}

Keeping the constellations and the transmit powers fixed, the achievable rates are a function of the strength of the interference and the rotation angle between the desired and the interfering part (ignoring the Gaussian noise) of the received signal. These rotation angles are
\begin{subequations}\label{eq:def_rho}
\begin{align}
 \rho_1 &= \psi+\varphi \label{eq:def_rho1}\\
 \rho_2 &= \psi-\varphi, \label{eq:def_rho2}
\end{align}
\end{subequations}
at receiver one and two, respectively. Hence, $\rho_1$ influences $I(Y_1;X_2)$, while $\rho_2$ influences $I(Y_2;X_1)$. Due to the symmetry of the QAM constellations and the Gaussian noise, it does not matter whether we take the exterior or interior angle between desired and interfering signal. Thus $\rho_i$ and $-\rho_i$ have the same effect at receiver $i$. This symmetry has an interesting consequence, which is formulated in the following proposition.\begin{prop}\label{lem:norotsym}
 In a 2-IC given by \reff{eq:2IC}, if no rotation is applied ($\varphi=0$), we have for $\alpha \in [0,2\pi]$
\begin{align}
 I(Y_1;X_2)\left|_{\psi=\phalf+\alpha} \right. &= I(Y_1;X_2)\left|_{\psi=\phalf-\alpha} \right. \\
 I(Y_2;X_1)\left|_{\psi=\phalf+\alpha} \right. &= I(Y_2;X_1)\left|_{\psi=\phalf-\alpha} \right. ,
\end{align}
i.e., $I(Y_1;X_2)$ takes on the same values for $\psi=\phalf+\alpha$ and $\psi=\phalf-\alpha$ and the same holds for $I(Y_2;X_1).$
\end{prop}
\begin{IEEEproof}
With no rotation, we have $\varphi = 0$ and thus 
\begin{equation}
\rho_1=\rho_2=\phalf + \alpha
\end{equation}
if $\psi=\phalf+\alpha$. For $\psi=\phalf-\alpha$, we have 
\begin{equation}
\rho_1=\rho_2=\phalf-\alpha=-\phalf-\alpha \mod \pper
\end{equation}
Since for the achievable rates with symmetric QAM constellation the sign of $\rho_i$ is irrelevant, both $I(Y_1;X_2)$ and $I(Y_2;X_1)$ take on the same value in both cases.
\end{IEEEproof}

Note that in the 2-Z-IC, where $I(Y_2;X_1)$ is always zero, Proposition \ref{lem:norotsym} also holds. Moreover, in the 2-Z-IC, the value of $\rho_2$ is pointless. Thus, finding the optimal rotation in the 2-Z-IC is much easier than in the 2-IC, because it influences only one term. Therefore, we now regard the problem of finding the optimal rotation for the 2-Z-IC and 2-IC separately in the next two subsections. For this optimization problem, we only regard those rate constraints that can be influenced by rotation, i.e., the sum rate constraints.

\subsection{2-Z-IC}

In the case of a 2-Z-IC \reff{eq:2ZIC} the only sum rate constraint is \reff{eq:mi_sum_Rx1}, which can be written as
\begin{equation}
 R_1+R_2 \leq \max_{\varphi}\;\; I(Y_1;X_2) + I(Y_1;X_1|X_2) \label{eq:2ZIC-sum}.
\end{equation}
As already stated, rotation does not influence the second term in \reff{eq:2ZIC-sum}, regardless whether this rotation is done by the receiver or the channel. The remaining term $I(Y_1;X_2)$ depends only on $\ah$ and $\rho_1$ given that the transmit power and $\X_1$, $\X_2$ are fixed. By choosing $\varphi$ at transmitter two, $\rho_1$ can be set to an arbitrary value between $0$ and $\pper$. Thus, any phase $\psi$ of the channel can be offset by rotation and does not influence the achievable rates.

\subsection{2-IC}
For the 2-IC, the situation is more complicated, since two constraints on the sum rate are active. They can be combined as
\begin{equation}
 R_1+R_2 \leq \max_{\varphi}\min \begin{Bmatrix}
I(Y_1;X_2) + I(Y_1;X_1|X_2),\\ I(Y_2;X_1) + I(Y_2;X_2|X_1)\end{Bmatrix} \label{eq:2IC-sum}.
\end{equation}
Again, $I(Y_1;X_1|X_2)$ and $I(Y_2;X_2|X_1)$ are not subject to rotation and due to the symmetry of the channel they are equal. Thus, for the 2-IC the optimal rotation is obtained by finding the maximum over $\varphi$ of the minimum of $I(Y_1;X_2)$ and $I(Y_2;X_1)$. These two terms depend on the phase shifts $\rho_1$ and $\rho_2$, respectively, which are (in general) unequal functions of $\varphi$ and $\psi$. Hence, the influence of a rotation $\varphi$ is different for the two crucial mutual information terms $I(Y_1;X_2)$ and $I(Y_2;X_1)$, whose minimum is additionally influenced by the channel phase. However, by rotation the influence of the channel phase can be mitigated, as the following proposition will show.

\begin{prop}\label{lem:rotsym}
Define
\begin{equation}
\I(\ah,\psi) = \max_\varphi \min \{ I(Y_1;X_2),I(Y_2;X_1)\} 
\end{equation}
as the minimum mutual information if the channel gains on the interference paths are $h=\ah\cdot\exp(i\psi)$. Then, the following symmetry relations hold ($\alpha,\beta \in [0,2\pi]$)
\begin{align}
 \I(\ah,\phalf+\alpha) &=  \I(\ah,\phalf-\alpha)\\
 \I(\ah,\pquart+\beta) &=  \I(\ah,\pquart-\beta)
\end{align}
\end{prop}
\begin{IEEEproof}
 The proof will be given in the appendix.
\end{IEEEproof}

\begin{figure}%
\centering
\subfloat[][]{\includegraphics[width=4cm]{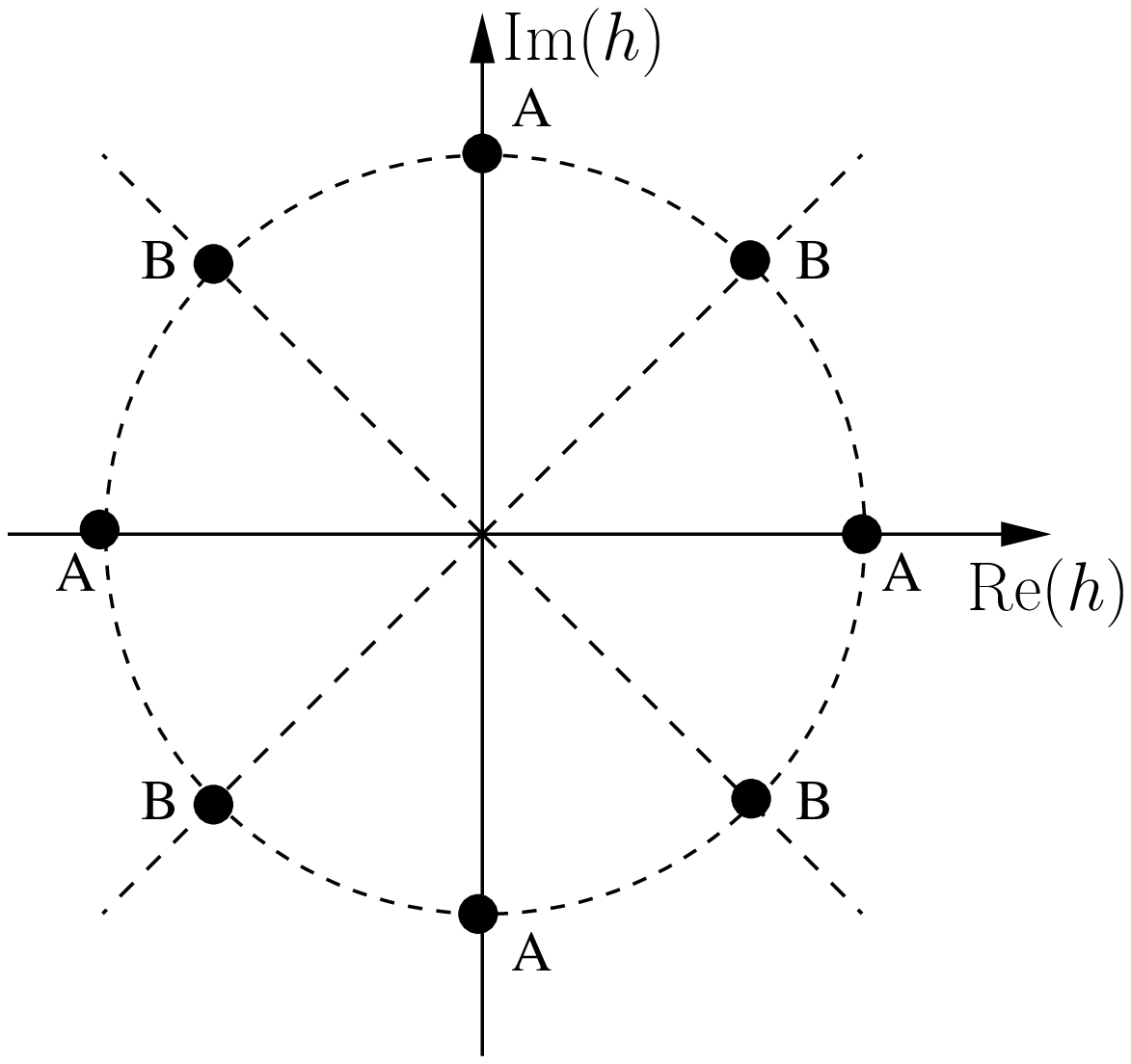}}%
\qquad
\subfloat[][]{\includegraphics[width=4cm]{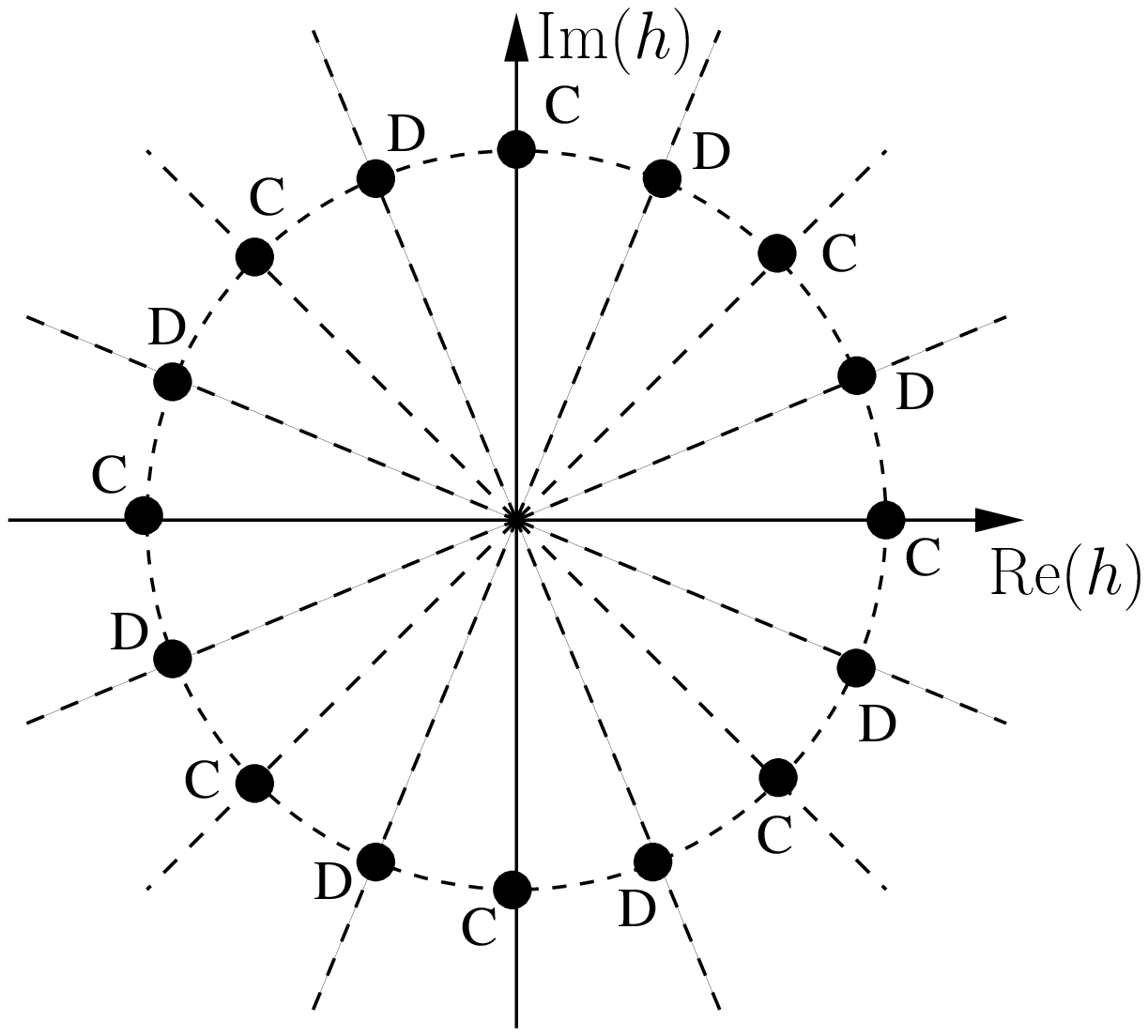}}
\caption{Illustration of Proposition \ref{lem:norotsym} (a) and Proposition \ref{lem:rotsym} (b) for QAM alphabets}%
\label{fig:visProp}%
\end{figure}

The Propositions \ref{lem:norotsym} and \ref{lem:rotsym} are illustrated in Figure \ref{fig:visProp} for the used QAM alphabets. In both parts of the figure, the absolute value $\ah$ of the channel gain is fixed, i.e., all possible values of $h$ lie on a circle. In the left part, we consider the situation without rotation. Proposition \ref{lem:norotsym} states that all values marked by ``A'' and ``B'' lead to the same performance, respectively. The same holds for the points ``C'' and ``D'' in the right part of the figure, where the situation with rotation is considered. It can be seen that with rotation, the points of equal performance get closer together. Moreover, both with and without rotation the dependence on $\psi$ is point-symmetric to all values of $\psi$ that correspond to a point marked by ``A'', ``B'', ``C'', or ``D''.

\section{Numerical Results}\label{sec:results}

In this section, we will evaluate the achievable sum rates of the 2-Z-IC and 2-IC by numerical simulations. Due to the different influence of the channel phase, we will discuss the 2-Z-IC and 2-IC separately. Subsequently, we will discuss the very strong interference threshold for the 2-IC.

\subsection{Z-IC}
\begin{figure}
  \includegraphics[width=9.5cm]{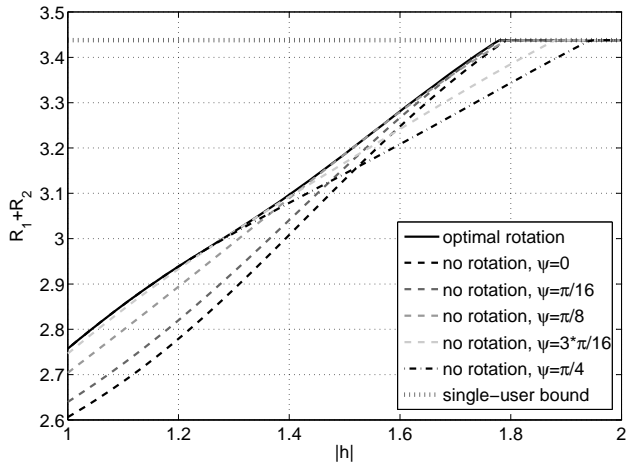}
  \vspace{-0.7cm}
  \caption{Achievable sum rates in 2-Z-IC, 4-QAM at $P=5$ dB}
  \vspace{-0.3cm}
  \label{fig:4-QAM_Zic}
\end{figure}

As stated in the previous section, the achievable sum rates in the 2-Z-IC are not subject to the channel phase $\psi$ if rotation is applied. However, this is not the case without rotation. The benefits of rotation can be seen in Figure \ref{fig:4-QAM_Zic}, where the achievable sum rates of 4-QAM are plotted for $P=5$ dB. It can be observed, that the achieved sum rate is always optimal if rotation is applied. Without rotation, the performance decreases and depends on the phase of the channel.

\subsection{2-IC}
\begin{figure}
 \includegraphics[width=9.5cm]{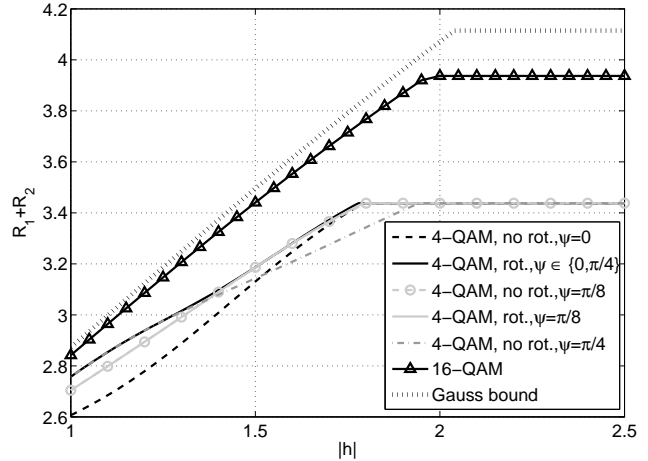}
  \vspace{-0.7cm}
 \caption{Achievable sum rates in 2-IC at $P=5$ dB}
  \vspace{-0.3cm}
 \label{fig:4-QAM_2IC}
\end{figure}

If no rotation is applied in the 2-IC, the achievable sum rates are the same as in the 2-Z-IC. This can be verified with \reff{eq:def_rho}, since $\varphi=0$ leads to $\rho_1=\rho_2$ and thus $I(Y_1;X_2)=I(Y_2;X_1)$. If rotation is applied, this does not hold anymore, since the influence on those terms is different, as stated in the preceding section. However, with rotation we can benefit from Proposition 2 and restrict our simulations to $\psi\in \left[ 0, \frac{\pper}{4}=\frac{\pi}{8} \right]$.

The achievable sum rates at $P=5$ dB with and without rotation are shown in Figure \ref{fig:4-QAM_2IC}. For 4-QAM, it can be seen that by rotation a substantial rate gain is obtained, especially for $\psi=0$ and $\ah$ close to $1$ as well as for $\psi=\frac{\pi}{4}$ and $\ah$ around $1.7$. Moreover, it can be observed that for $\psi=\pi/8$, rotation can not increase the performance significantly. Another important observation is that for certain cases (e.g. $\psi=\pi/4$), rotation has the effect that the system is pushed in the very strong interference regime.

If 16-QAM is used for transmission, we found that the influence of rotation (by transmitter or channel) is marginal for this value of $P$. For clarity of the Figure, we plotted only one line for 16-QAM (the variations due to rotation are within $\pm10^{-2}$). Note, that this is also the performance of 16-QAM in the 2-Z-IC. However, for larger values of $P$, we can see the same influence of rotation as observed for 4-QAM at $P=5$ dB. For comparison, we also added the capacity region achieved by Gaussian input alphabets to Figure \ref{fig:4-QAM_2IC}.

\subsection{Threshold between strong and very strong interference}
\begin{figure}
 \includegraphics[width=9.5cm]{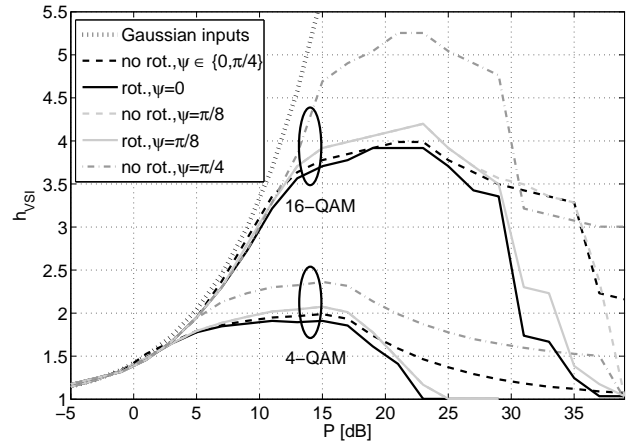}
  \vspace{-0.7cm}
 \caption{Very strong interference threshold $\hV$ for 2-IC}
  \vspace{-0.3cm}
 \label{fig:VSI}
\end{figure}

As already stated in subsection \ref{subsec:capreg}, interference does not reduce capacity, i.e., we are in the very strong interference regime, if $\ah$ is bigger than a certain threshold $\hV$. For Gaussian alphabets, this value is known to be $\hV = \sqrt{1+P}$. From Figure \ref{fig:4-QAM_2IC} it is observed that the threshold $\hV$ is lower if finite alphabets are used. The exact value depends on the alphabet used and the rotations by transmitter and channel. However, Figure \ref{fig:4-QAM_2IC} considers only the case $P=5$ dB. In this subsection, we will analyze how $\hV$ varies with the available transmit power $P$.

In Figure \ref{fig:VSI}, $\hV$ is plotted for different values of $P$ both with rotation (solid lines) and without (dashed lines). It can be seen that $\hV$ is always lower than $\sqrt{1+P}$ if finite alphabets are used. Moreover, it can be seen that $\hV$ converges to 1 for both small and large values of $P$. For small values of $P$ it was already observed in \cite{Ver02} that complex finite alphabets behave like Gaussian alphabets, which is consistent with our observations. For large values of $P$, the explanation is as follows: As $P\rightarrow \infty$ and $h>1$, the channel can be considered as noiseless. For this case, it can be shown that the constraints \reff{eq:mi_sum_Rx1} and \reff{eq:mi_sum_Rx2} are obtained from \reff{eq:mi_R1} and \reff{eq:mi_R2} by addition. Thus, the strong interference regime vanishes.

Also in Figure \ref{fig:VSI}, the benefits of optimizing the rotation can be observed. Especially for the unfavorable case of $\psi=\pi/4$, rotation can decrease $\hV$. This means that by rotation, the channel can be pushed in the very strong interference regime. As already stated in the previous subsection, for 16-QAM the benefits of rotation become visible for values of $P$ above around $12$ dB, while with 4-QAM the benefits of rotation are already obtained for lower values of $P$. For 4-QAM it can be observed again that at $\psi=\pi/8$, rotation can not increase the performance significantly. However, for 16-QAM this is not the case. Instead, we can observe that if $P$ is around $30$ dB, rotation can help to decrease $\hV$.

\section{Conclusion}\label{sec:conclusion}
In this paper, we have analyzed the achievable rates in the strong and very strong interference regime of two-user interference channels. Although it is well-known, that the capacity region is achieved by Gaussian alphabets, these alphabets are continuous and unbounded, which makes them unsuitable for practical systems. Therefore, we restricted the inputs to finite QAM-constellations, that are used in real world applications.

We have shown that the performance is subject to the phases of the channel gains. However, the influence of these phases can be mitigated if the transmit alphabets are pre-rotated at one of the transmitters. Moreover, by this rotation, we have a degree of freedom which can be used to optimize the achievable rates. It was observed by numerical simulations that the rate increase by rotation is significant for a large set of parameters. Finally, we analyzed the threshold between the strong and very strong interference regime for the considered finite alphabets. It could be seen that this threshold is always lower than for Gaussian alphabets and can be optimized by rotation, too.

\section*{Acknowledgment}
This work was supported by the German research council ``Deutsche Forschungsgemeinschaft'' (DFG) under grants Bo 867/18-1 and Se 1697/3-1.

\appendix
\subsection{Proof of Proposition \ref{lem:rotsym}}

\begin{IEEEproof}
First, we prove the first statement of the proposition. Suppose that for $\psi=\phalf+\alpha$, the optimal rotation is given by $\varphi=\popt$. Thus, we have 
\begin{subequations}
\begin{align}
\rho_1 &= \phalf+\alpha+\popt \\
\rho_2 &= \phalf+\alpha-\popt.
\end{align}
\end{subequations}
Now, for $\psi=\phalf-\alpha$, let us choose $\varphi =-\popt$. Then, we obtain
\begin{subequations}
\begin{align}
\rho_1 &= \phalf-\alpha-\popt = -\phalf-\alpha-\popt \mod \pper\\
\rho_2 &= \phalf-\alpha+\popt = -\phalf-\alpha+\popt \mod \pper
\end{align}
\end{subequations}
Since these interference phase shifts only differ from those obtained for $\psi=\phalf+\alpha$ by their sign, and due to the symmetry of the QAM constellations, we can conclude that the values of both $I(Y_1;X_2)$ and $I(Y_2;X_1)$ are the same for both angles.

A similar technique can be applied for proving the second statement. For a channel phase of $\psi=\pquart+\beta$, assume the optimal rotation is given by $\varphi=\popt$. Then, for the case of $\psi=\pquart-\beta$ we choose $\varphi=\phalf-\popt$, which results in the same $\rho_1$ and $\rho_2$ as for $\psi=\pquart+\beta$. The calculation and argumentation is in analogy to the proof of the first part.
\end{IEEEproof}



%

\bibliographystyle{IEEEtran}

\end{document}